\documentclass[12pt,preprint]{aastex}

\usepackage{multirow}

\def\um{\micron}

\def\gapp{\lower 3pt\hbox{${\buildrel > \over \sim}$}\ }
\def\lapp{\lower 3pt\hbox{${\buildrel < \over \sim}$}\ }
\def\proptosim{\lower 3pt\hbox{${\buildrel \propto \over \sim}$}\ }

\def\arcsec{$^{\prime\prime}$}
\def\degree{$^\circ$}

\begin{document}

\title{Determining the Shape, Size, and Sources of the Zodiacal Dust Cloud \\
       Using Polarized Ultraviolet Scattered Sunlight}

\author[0000-0001-5966-837X]{Geoffrey Bryden}
\affil{Jet Propulsion Laboratory, California Institute of Technology, 4800 Oak Grove Dr., Pasadena, CA 91109, USA}

\author[0000-0001-8292-1943]{Neal J. Turner}
\affil{Jet Propulsion Laboratory, California Institute of Technology, 4800 Oak Grove Dr., Pasadena, CA 91109, USA}

\author[0000-0002-5667-9337]{Petr Pokorn\'{y}}
\affil{Department of Physics, The Catholic University of America, Washington, DC 20064, USA}
\affil{Astrophysics Science Division, NASA Goddard Space Flight Center, Greenbelt, MD 20771, USA}
\affil{Center for Research and Exploration in Space Science and Technology, NASA/GSFC, Greenbelt, MD 20771, USA}

\author[0000-0001-8292-1943]{Youngmin Seo}
\affil{Jet Propulsion Laboratory, California Institute of Technology, 4800 Oak Grove Dr., Pasadena, CA 91109, USA}

\author[0000-0001-8292-1943]{Brian Sutin}
\affil{Jet Propulsion Laboratory, California Institute of Technology, 4800 Oak Grove Dr., Pasadena, CA 91109, USA}

\author[0000-0001-6403-841X]{Virginie Faramaz}
\affil{Steward Observatory, University of Arizona, Tucson, AZ 85721, USA}

\author{Keith Grogan}
\affil{Jet Propulsion Laboratory, California Institute of Technology, 4800 Oak Grove Dr., Pasadena, CA 91109, USA}

\author[0000-0002-0435-8224]{Amanda Hendrix}
\affil{Planetary Science Institute, Tucson, AZ 85719 USA}

\author[0000-0003-4205-4800]{Bertrand Mennesson}
\affil{Jet Propulsion Laboratory, California Institute of Technology, 4800 Oak Grove Dr., Pasadena, CA 91109, USA}

\author{Susan Terebey}
\affil{California State University Los Angeles, Los Angeles, CA 90032, USA}

\begin{abstract}
  The solar system's Zodiacal Cloud is visible to the unaided eye, yet the origin of its constituent dust particles is not well understood, with a wide range of proposed divisions between sources in the asteroid belt and Jupiter Family comets.  The amount of dust contributed by Oort Cloud comets is uncertain.
  Knowledge of the Zodiacal Cloud's structure and origins would help with NASA's aim of characterizing potentially Earth-like planets around nearby stars, since the exo-Earths must be studied against the light scattered from extrasolar analogs of our cloud. 
  As the only example where the parent bodies can be tracked, our own cloud is critical for learning how planetary system architecture governs the interplanetary dust's distribution. 
  Our cloud has been relatively little-studied in the near-ultraviolet, a wavelength range that is important for identifying potentially-habitable planets since it contains the broad Hartley absorption band of ozone.
  We show through radiative transfer modeling that our cloud's shape and size at near-UV wavelengths can be measured from Earth orbit by mapping the zodiacal light's flux and linear polarization across the sky.     
  We quantify how well the cloud's geometric and optical properties can be retrieved from a set of simulated disk observations, using a Markov chain Monte Carlo analysis.
  The results demonstrate that observations with sufficient precision, covering a set of fields distributed along the ecliptic and up to the poles, can be used to determine the division between asteroidal, Jupiter Family, and Oort Cloud dust components, primarily via their differing orbital inclination distributions.
  We find that the observations must be repeated over a time span of several months in order to disentangle the zodiacal light from the Galactic background using the Milky Way's rotation across the sky.
\end{abstract}

\submitjournal{PASP}
\shorttitle{Polarized UV Observations of Zodiacal Dust}
\shortauthors{Bryden et al.}


\newpage
\section{Introduction}

The Zodiacal Cloud (ZC) is a continuously replenished structure composed of dust grains and meteoroids, stretching from a few solar radii from the Sun \citep{stenborg21} to locations beyond the Kuiper belt \citep{poppe19}.
Due to radiative \citep{burns79} and collisional processes \citep{gruen85} that remove dust on relatively short ($<$Myr) timescales, the Zodiacal Cloud presents a recent imprint of the dust generating activity of various source populations. 
It is continuously evolving as dust particles and meteoroids are released through collisions among and sublimation of their asteroidal and cometary parent bodies, and subsequent collisions between the dust particles and meteoroids themselves.

The major reservoirs of source bodies in the solar system range from
main-belt asteroids and Jupiter-family comets (JFCs) in the inner solar system to
Halley-type and Oort-Cloud comets, and Kuiper-belt objects \citep{nesvorny11, Pokorny2014}.
As the locations and inclinations of these reservoirs imprint different spatial characteristics on dust within the ZC, the dust carries crucial information about the relative abundances of parent bodies and the evolution of the Zodiacal Cloud \citep{hahn02}.

The distinct dynamical imprints in the ZC from the separate sources arise from their different dynamical histories. Dust grains are released with orbital elements similar to their parent bodies, which vary significantly between source populations that are concentrated toward the ecliptic plane (asteroids and Jupiter-family comets) and those that follow an isotropic distribution (Oort cloud comets).
Once released, in addition to solar and planetary gravity, dust grains are also influenced by solar radiation, the solar wind, and the solar system magnetic field \citep{burns79}. The radiative forces act on dust via two different effects: (1) radiation pressure pushes the particles radially away from the Sun, counteracting the Sun's gravity enough to blow out the smallest grains from the system, and (2) Poynting-Robertson (PR) drag effectively pushes against the particles' motion, causing their orbits to spiral in towards the Sun. 
While interplanetary electric and magnetic fields are important for the dynamics of nanometer-scale particles \citep{czechowski10}, they are a minor force on the grains larger than a few microns which provide most of the optical cross-section of the ZC.

While the smallest grains ($\lesssim$1~\um, depending on composition) can be quickly blown out of the solar system,
larger grains remain orbiting the Sun for hundreds to millions of years, undergoing often-complex dynamical evolution.  Every particle born on a bound orbit is eventually removed by either: (1) evaporation close to the Sun after spiraling down under the influence of PR drag, (2) collision with a planet or large asteroid, (3) ejection from the solar system after a close encounter with a planet, or (4) destruction in a collision with another particle. 
Smaller grains have shorter lifetimes under PR drag, while larger particles ($\gtrsim$100 \um) are likely to be fragmented in collisions with other particles before they spiral down very far \citep{gruen85, wyatt05}.  Hence the Zodiacal Cloud needs to be continuously replenished and the sources of its resupply leave imprints in the orbits of the particles making up the cloud.

The sources of dust in the inner solar system -- main-belt asteroids and Jupiter-family comets -- produce components of the ZC that are close to the ecliptic, arriving at Earth with both eccentricity and inclination low \citep{nesvorny10, nesvorny11, Pokorny2019, Koschny2019}. 
The sources of dust originating in the outer solar system, on the other hand, form a more isotropic component of the ZC, with orbital inclinations ranging from 0\degree\ to 180\degree\ and relatively high eccentricities \citep{nesvorny11, Pokorny2014}.  There are two general reasons for this dichotomy.  The first is simply that each grain is created with an inclination similar to its parent body, and the inner solar system source populations (the Jupiter-family comets and asteroids) are generally on low-inclination orbits confined within 30\degree\ of the ecliptic \citep{nesvorny17}, whereas the outer solar system source populations include the broad scattered disk and the Oort Cloud with their higher inclinations \citep{vokrouhlicky19}.  Second, dust grains originating from inner solar system sources do not need to overcome Jupiter’s gravitational barrier, while outer solar system particles must pass by Jupiter to enter the inner solar system and are often dynamically scattered (and even ejected) in the process.  The Jupiter barrier effectively protects the inner solar system from Kuiper-belt particles \citep{poppe16}.

Knowing the relative abundances of dust from the different populations of source bodies would be a substantial advance in our understanding of the ZC's structure and evolution, yet these abundances remain poorly constrained by existing observations.
While models of asteroid collisions matched to the IRAS dust bands find that the majority of zodiacal dust is asteroidal in origin \citep{dermott02},
an analysis of cometary dust production and dynamical evolution yields the opposite conclusion, that most dust comes from Jupiter-family comets \citep{nesvorny10}.
The same analysis finds that up to 10\% of the dust could be from long-period comets, a possibility supported by several recent observations suggesting nearly-isotropic components in the optical and infrared sky backgrounds \citep{Sano2020, Lauer2021, Korngut2022}.
      
We consider in this paper whether new observations from a dedicated space-based telescope can resolve
this issue, determining the relative fractions of asteroidal vs.\ cometary dust and detecting for the first time the dust from Oort Cloud comets, should it exist as a significant component.
Fortunately the Earth’s location at 1~au is in a critical ``sweet spot'' for probing the dynamics of different ZC populations and revealing the sources and structure of the ZC; all major ZC source populations have semi-major axes larger than 1~au, which allows detection of meteoroids from all sources in different stages of dynamical evolution.  Meanwhile, the Earth is far enough from the Sun that many particles survive the journey to 1~au without being destroyed in collisions, unlike the particles impacting Mercury \citep{gruen85, pokorny18}. 

In the following, we first develop a model for the sky brightness that includes contributions from both the Zodiacal Cloud (\S\ref{sec:zcmodel}) and background Galactic sources (\S\ref{sec:galBack}).
We use this sky model to generate synthetic observations as might be obtained with a small Earth-orbiting satellite (\S\ref{observations}).
We then apply Markov-chain Monte Carlo analysis to the observations to retrieve the Zodiacal Cloud's parameters, as a function of qualities of the observations including the signal-to-noise ratio, number of fields observed, and number of epochs (\S\ref{results}).  Finally, we summarize the findings in (\S\ref{summary}).

\section{Zodiacal Cloud Model\label{sec:zcmodel}}

We consider a model for the Zodiacal Cloud that is 
matched to current observational constraints.
We do not trace its physical evolution -- dust creation through asteroid collisions and comet sublimation; subsequent orbital migration -- instead adopting a set of phenomenological parameters to describe the disk structure and its radiative properties.

\subsection{Spatial Distribution}\label{dustDistribution}

\begin{figure}\begin{center}
  \includegraphics[width=3.5in]{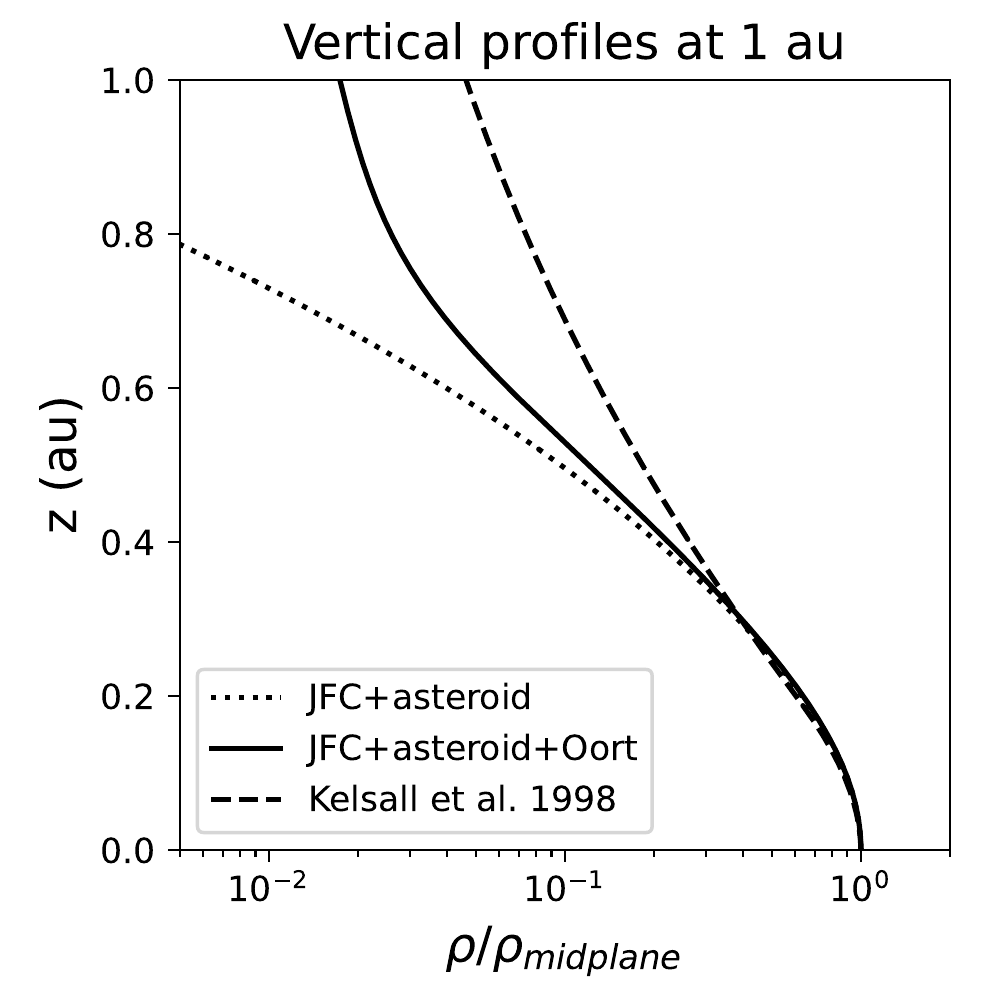}
  \end{center}
  \caption{
  Dust from asteroids and Jupiter-family comets is concentrated in the disk midplane, following Gaussian distributions with H/R = 0.15 and 0.25 for the two components (their combined vertical profile is shown as a dotted line).
  Including dust from Oort Cloud comets (solid line)
  increases the dust density away from the midplane.
  This three-component model for the dust is more consistent with \citet{kelsall98} profile fit to COBE's all-sky map of thermal emission.
  }\label{vertProfiles}
\end{figure}

Our geometric model of the disk is based on an all-sky map of the dust's 
thermal emission observed by the Cosmic Background Explorer (COBE) mission's
Diffuse Infrared Background Experiment (DIRBE) infrared instrument
\citep{Silverberg1993}. 
\citet{kelsall98} fit the zodiacal dust component of this COBE/DIRBE map 
across wavelengths from 1.25 to 240 \um, constraining the overall shape of the Zodiacal cloud. 
While the Kelsall model provides detailed measurement of some disk substructure 
including three dust bands and an Earth-trailing blob of enhanced emission,
it does not distinguish between the different physical components of the disk.
In particular, \citet{kelsall98} do not provide any explicit constraints on a spherical component to the disk (as would be supplied by Oort cloud comets).
Instead they model the vertical disk profile as a 3-parameter piece-wise exponential 
(dashed line in Figure~\ref{vertProfiles}).

We consider a disk model with three components, each with its own vertical profile.
The asteroidal and Jupiter-family comet components are modelled as Gaussians
with scale heights of 0.15 and 0.25 times the radius, respectively, 
matching the distribution of dust produced by the known asteroids and Jupiter-family comets,
which have average inclinations of 9.0\degree\ and 13.5\degree\ respectively
\citep{Moskovitz2022}.
The dust from these two components is concentrated toward the ecliptic plane,
much more so than the gradual dropoff with height seen in the Kelsall model
(dotted vs.\ dashed lines in Figure \ref{vertProfiles}).
For consistency with the DIRBE observations -- and because Oort Cloud dust must exist at some level -- we include a third, spherically symmetric dust component.
For the fraction of dust in each component we adopt a 70/20/10 split in surface density between Jupiter family comets, asteroids, and Oort cloud comets,
with the total surface density at 1 au set to match the \citet{kelsall98} optical depth of $8\times 10^{-8}$.
While the combined vertical profile (solid line in Figure \ref{vertProfiles}) is a better match to the Kelsall model, additional Oort Cloud dust would be needed for a perfect match.
Instead we limit the model to a 10\% fraction of Oort Cloud dust as a conservative guess for this key unknown parameter.
Note that the split between the three components is in terms of vertically integrated optical depth;
the Oort Cloud dust only contributes 4\% of the volume density in the disk midplane
(5.6$\times 10^{-9}$ au$^{-1}$, cf.\ the total midplane density of 1.4$\times 10^{-7}$ au$^{-1}$, 
in terms of cross sectional area per volume).

For the radial distribution of the dust, we again follow \citet{kelsall98}, 
who find a gradually declining dust surface density moving away from the Sun
(proportional to orbital radius to the $-$0.34 power).
For the spherically-symmetric Oort-Cloud component, we assume a steeper falloff in volume density $\rho \propto r^{-2.5}$.
We apply an outer cutoff at 3.3~au, as observed by in-situ photometry from the Pioneer 10 and 11 spacecraft as they traveled to the outer Solar System \citep{Mann2006}.

The overall density distribution in our model is then
\begin{equation}
\rho_{\rm ZC}(R, z) = 
\tau_{\rm asteroid} \, R^{\gamma_{\rm asteroid}} \, e^{-\frac{1}{2}(z/H_{\rm asteroid})^2}
+
\tau_{\rm JFC} \, R^{\gamma_{\rm JFC}} \, e^{-\frac{1}{2}(z/H_{\rm JFC})^2}
+
\rho_{\rm Oort} \, R^{\gamma_{\rm Oort}} 
\,\,\,\, [\rm{for} \,\, R < R_{\rm out}]
\end{equation}
with the parameters described and the default values summarized in 
Table~\ref{diskParamTable}.

Note that in addition to the nine geometric parameters prescribing the dust density distribution,
there are three additional parameters describe the dust scattering properties (\S\ref{dustscat})
and another four to fit the smooth part of the Galactic background (\S\ref{sec:galBack}).
Each model thus has a total of 16~parameters to be determined from the synthetic observations as described in (\S\ref{observations}).

\begin{deluxetable}{cc|c|l}
\tablecaption{Zodiacal Cloud Model Parameters
\label{diskParamTable}}
\tablehead{Parameter & Description & Value & Motivation }
\startdata
$R_{\rm out}$ & outer edge & 3.3 au & Pioneer 10,11 in-situ photometry \citep{Mann2006} \\
\multicolumn{4}{c}{Asteroidal component} \\
\hline
$\tau_{\rm asteroid}$ & optical depth & $1.6 \times 10^{-8}$ & 
20\% of $\tau_{\rm total}$ \citep{kelsall98} \\
$\gamma_{\rm asteroid}$ & $\tau$ radial exponent & $-$0.34 &
COBE/DIRBE thermal emission \citep{kelsall98} \\
$H/R_{\rm asteroid}$ & disk height ratio & 0.15 &
known asteroid distribution \citep{Moskovitz2022} \\
\multicolumn{4}{c}{Jupiter-family comet component} \\
\hline
$\tau_{\rm JFC} $ & optical depth & $5.6 \times 10^{-8}$ &
70\% of $\tau_{\rm total}$ \citep{kelsall98} \\
$\gamma_{\rm JFC}$ & $\tau$ radial exponent  & $-$0.34 &
COBE/DIRBE thermal emission \citep{kelsall98} \\
$H/R_{\rm JFC}$ & disk height ratio & 0.25 & 
known comet distribution \citep{Moskovitz2022} \\
\multicolumn{4}{c}{Oort-cloud comet component} \\
\hline
$\rho_{\rm Oort}$ & density at 1 au & 0.56 $\times 10^{-8}$ &
10\% of $\tau_{\rm total}$ \citep{kelsall98} \\
$\gamma_{\rm Oort}$ & $\rho_{\rm Oort}$ radial exponent & $-$2.5 & sublimation-based estimate \\
\hline
\enddata
\end{deluxetable}

\subsection{Dust Scattering Properties}\label{dustscat}

The focus of this paper is the sunlight scattered and polarized by zodiacal dust at near-ultraviolet wavelengths around 280~nm;
thermal emission from the dust at infrared wavelengths is not considered.
Because the ZC is optically-thin, calculating the scattered light flux as observed from  Earth orbit is a straightforward integration along each chosen line-of-sight.
To calculate the amount and direction of polarization, we keep track of the light scattered parallel and perpendicular to the scattering plane containing the Sun, dust particle, and observer.

For the scattered light's angular dependence we use the functional form that \citet{hong85} fit to the zodiacal light at optical wavelengths.
The zodiacal dust is both strongly forward-scattering and significantly back-scattering, so it is not well-approximated by a single Henyey-Greenstein scattering phase function,
\begin{equation}
  \Phi(\theta) = \frac{1 - g^2}{[1 + g^2 - 2 g \cos(\theta)]^{3/2}}
\end{equation}
where $\theta$ is the scattering angle and $g\in\left[-1,1\right]$ governs the anisotropy, ranging from $-1$ for complete back-scattering to $+1$ for complete forward-scattering \citep{1941ApJ....93...70H}.
Instead, Hong fit the scattered intensity versus angle with a sum of three Henyey-Greenstein scattering functions 
(Table~\ref{dustProps}).
In modeling below, we fit the overall dust albedo and the degree of forward scattering ($g$ for the largest-weight Hong component) for the  
scattering of radiation polarized parallel and perpendicular to the scattering plane.
The remaining scattering parameters are fixed at the values \citet{hong85} determined by fitting the optical zodiacal light.

\begin{deluxetable}{cc|cc}
  \tablecaption{Assumed Dust Scattering Properties    
  \label{dustProps}}
  \tablehead{
    \shortstack{Polarization\\Direction} & Albedo &  $g$ 
    & \hskip 0.1in Weight \hskip 0.1in 
  }
  \startdata
Perpendicular & 0.43 & $+$0.68    & 0.636 \\
              &      & $-$0.14 & 0.361 \\
              &      & $-$0.84 & 0.003 \\
\hline
Parallel      & 0.34 & $+$0.72    & 0.694 \\
              &      & $-$0.28 & 0.299 \\
              &      & $-$0.81 & 0.006 \\
\enddata
\tablecomments{
Adopted from the \citet{hong85} fit to the Zodiacal Cloud.}
\end{deluxetable}

\begin{figure}\begin{center}
  \includegraphics[width=3.5in]{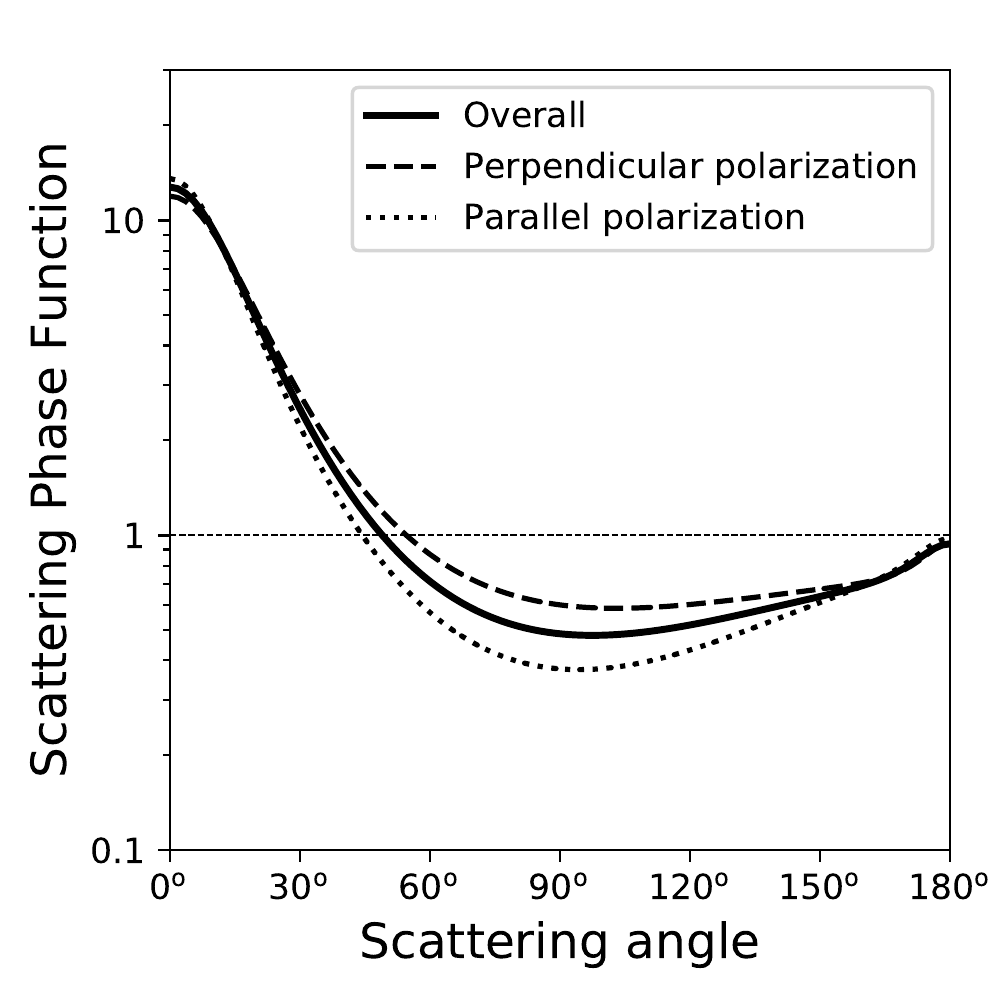}
  \includegraphics[width=3.5in]{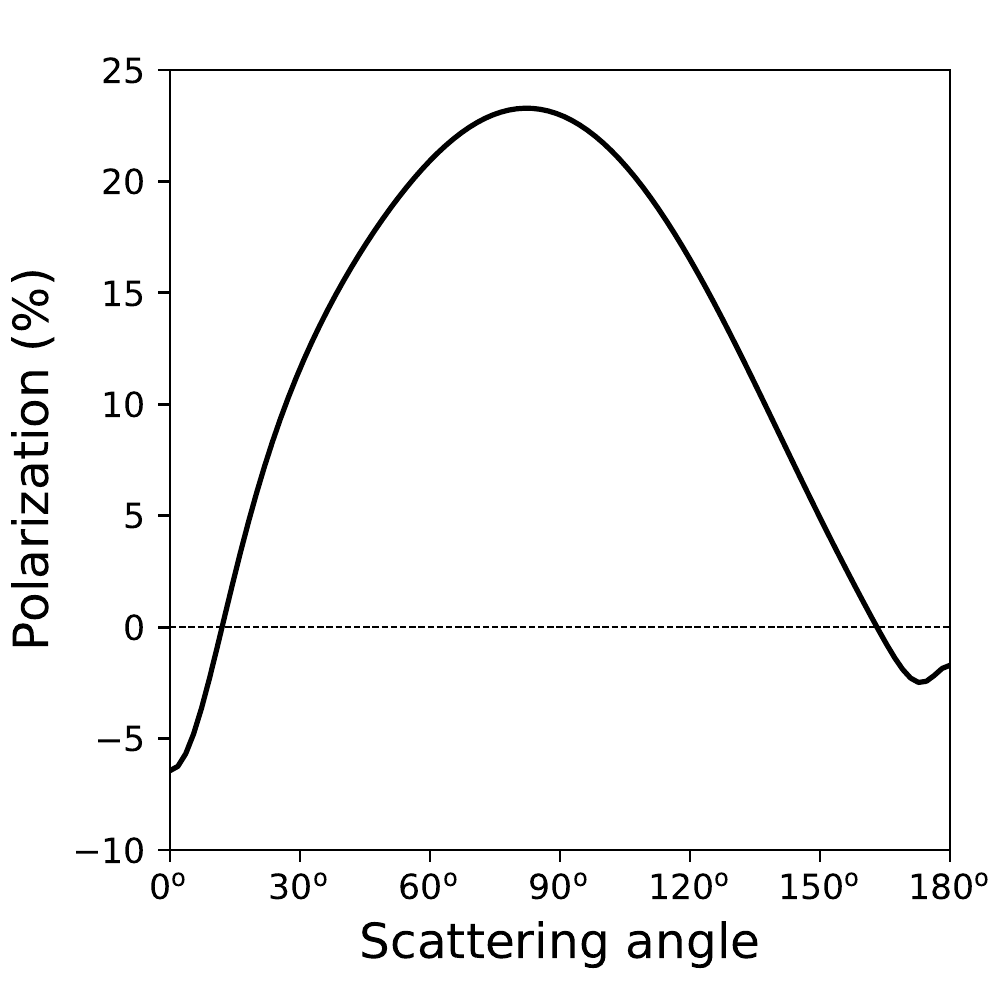}
  \end{center}
  \caption{
    {\it Left:}
    The net polarization fraction for zodiacal dust is shown
    as a function of scattering angle, based on the \citet{hong85} fit to the Zodiacal Cloud at optical wavelengths.
    At $\sim$90\degree\ scattering angles, light is preferentially scattered with polarizations perpendicular to the scattering plane,
    while the reverse is true close to 0\degree\ and 180\degree.
    {\it Right:}
    Scattering at a right angle produces strong ($>$20\%) polarization
    perpendicular to the scattering plane.
    The polarization is weaker for nearly forward or backward scattered light;
    the negative values there indicate polarization parallel to the scattering plane.
  }\label{polarization}
\end{figure}

Figure \ref{polarization} shows the scattering phase function and the
net polarization produced by these assumed dust properties.
The difference between the parallel and perpendicular scattering efficiencies results in a net linear polarization, but the direction and magnitude of this polarization varies with scattering angle.
For $\sim$90\degree\ scattering angles, light is preferentially scattered with polarizations perpendicular to the scattering plane (dashed line in the left panel of Figure \ref{polarization}), resulting in strong net polarization (more than 20\%).
Looking closer to or farther away from the Sun,
the polarization is both weaker and rotated;
the negative values in the right panel of Figure \ref{polarization} indicate polarization parallel to the scattering plane, rather than perpendicular. 

\begin{figure}\begin{center}
 \includegraphics[width=4.5in]{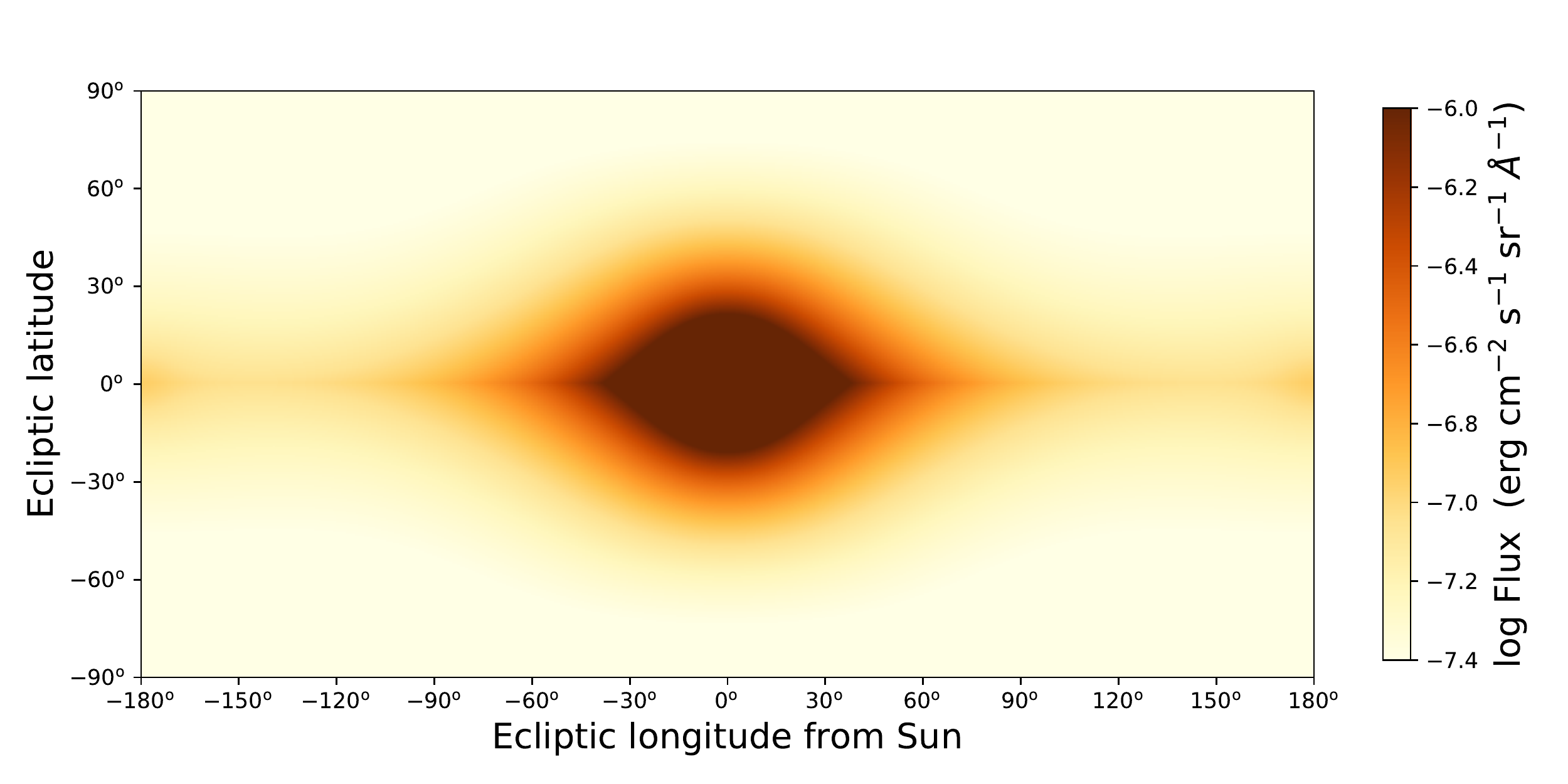}
 \includegraphics[width=4.5in]{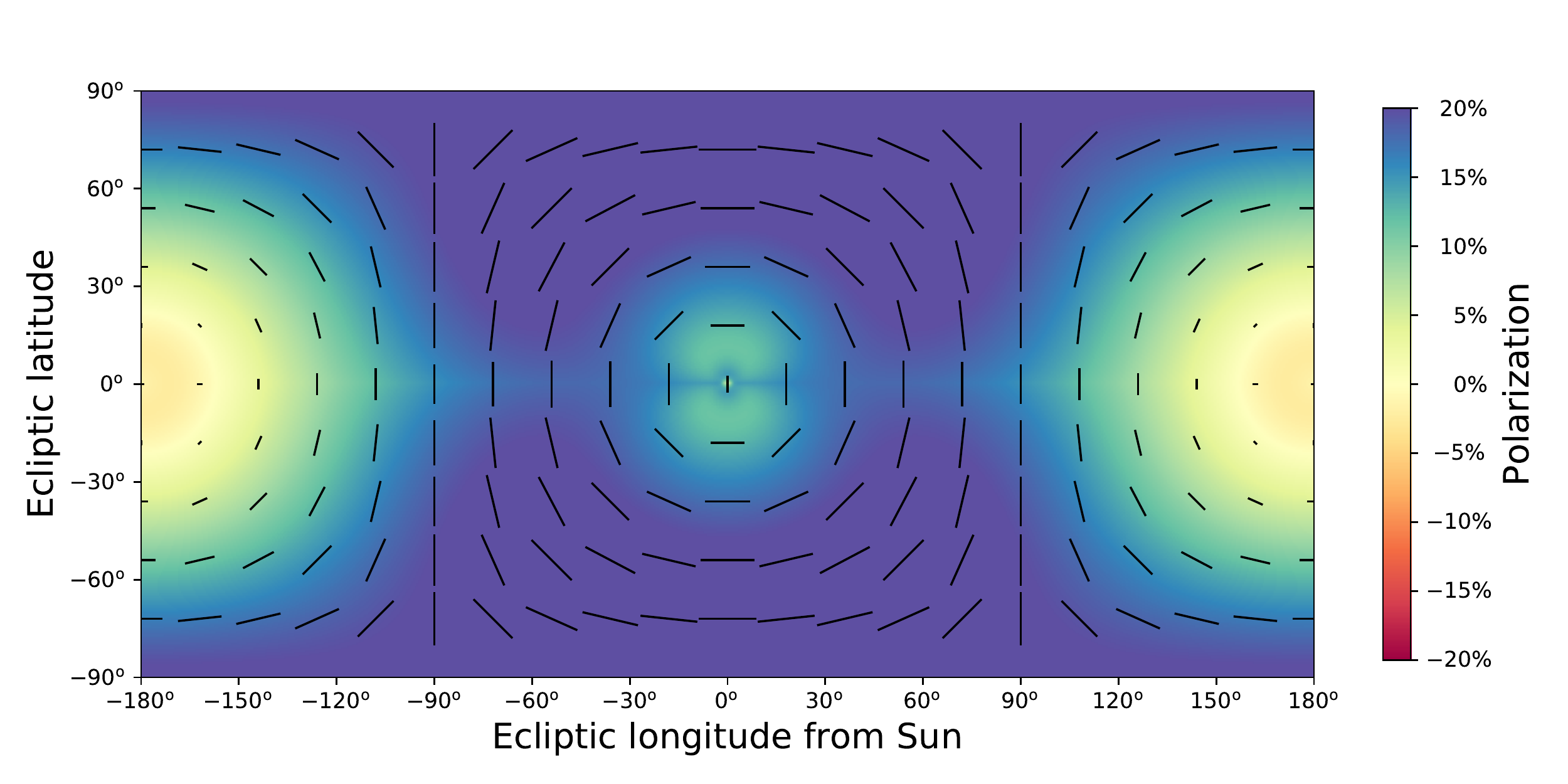}
  \end{center}
  \caption{Total intensity (top) and polarization (bottom) within a near-UV bandpass,
  for a model with just the zodiacal light and no Galactic contamination.
  The polarization vectors (black lines) are concentric around the Sun,
  with expected polarization fraction ranging up to nearly 20\% for light scattered at $\sim$90\degree\ angles.
  }\label{zodionly}
\end{figure}

Figure \ref{zodionly} shows the flux and polarization for our ZC model alone, without the background Milky Way and without any instrumental noise.
The zodiacal flux varies smoothly across the sky, with emission in the ecliptic plane a factor of $\sim$3-4 brighter than the ecliptic pole, except where it rises dramatically close to the Sun, consistent with published flux tables as a function of ecliptic angle \citep{leinert98}.
The polarization map has more structure, varying across the sky in both magnitude and direction, from a peak of 18\% polarization when viewing at angles roughly perpendicular to the Sun down to $-$2\% polarization (negative value indicates polarization parallel to the scattering plane, rather than perpendicular) when pointing nearly away from the Sun.

\section{Galactic Background}\label{sec:galBack}

\begin{figure}\begin{center}
  \includegraphics[width=4.5in]{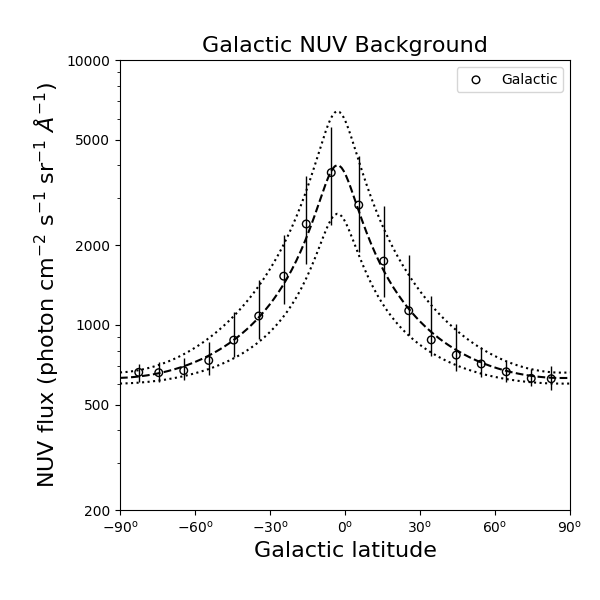}
  \end{center}
  \caption{GALEX observations of the NUV Galactic background (open circles, with error bars showing the dispersion between pointings) find much higher emission toward the Galactic midplane, where star-forming regions are concentrated.
  While the overall trend can be fit by a modified cosecant (dashed line; Equation \ref{eq:galBack}) the emission varies significantly at each latitude.
  }\label{fig:galBack}
\end{figure}

Emission from the Zodiacal Cloud can be contaminated by other sources in the sky, most significantly by the background of Galactic stars and interstellar dust.  We include this source of noise in our models based on measurements from the Galaxy Evolution Explorer (GALEX) of the Galactic background at 231~nm, close to the 280-nm wavelength of interest here.
Although GALEX was a pointed 
mission, its large (1.2\degree\ diameter) field of view enables the creation of nearly all-sky UV maps at resolution as fine as 6\arcmin\ \citep{Murthy2010, Hamden2013, Murthy2014}.
\citet{Murthy2014} also provide a catalog of GALEX's 44,843 observations with $\sim$233 million lines of individual measurements across the sky.
At each of these locations they provide the FUV (153~nm) and NUV (231~nm) flux of the diffuse Galactic background. 
Binning these data by Galactic latitude yields an overall trend of higher emission in the Galactic plane, roughly following a cosecant law \citep{Murthy2010}.
To obtain a model for the Galactic emission that can be subtracted from the observations to determine the zodiacal light contribution, we approximate the GALEX-measured trend with a cosecant offset to the south and also apply a maximum so the cosecant does not reach infinity in the Galactic plane:
\begin{equation}
F_{Galactic}(\ell) \approx F_0 + F_{\rm slope}     
    \left(\frac{\sqrt{C_0^2 + 1} }
         {\sqrt{C_0^2 + \rm{sin}^2(\ell-\ell_{\rm offset})} } - 1
         \right)\label{eq:galBack}
\end{equation}
where $\ell$ is the Galactic latitude,
$F_0$ = 630 photon cm$^{-2}$ s$^{-1}$ Sr$^{-1}$ \AA$^{-1}$ is the minimum at the Galactic poles,
$F_{\rm slope}$ = 500 photon cm$^{-2}$ s$^{-1}$ Sr$^{-1}$ \AA$^{-1}$ is the slope, 
$\ell_{\rm offset}$ is a 3\degree\ shift of the peak to the south, and
$C_0$ = 0.13 adds a cap to the peak emission.
Note that this flux is considerably higher than \citet{Murthy2014}
because we are fitting the median flux, rather than calculating a minimal floor to the background.
Figure \ref{fig:galBack} shows the binned GALEX data as points, while the dashed line shows our fit.

\begin{figure}\begin{center}
  \includegraphics[width=3.5in]{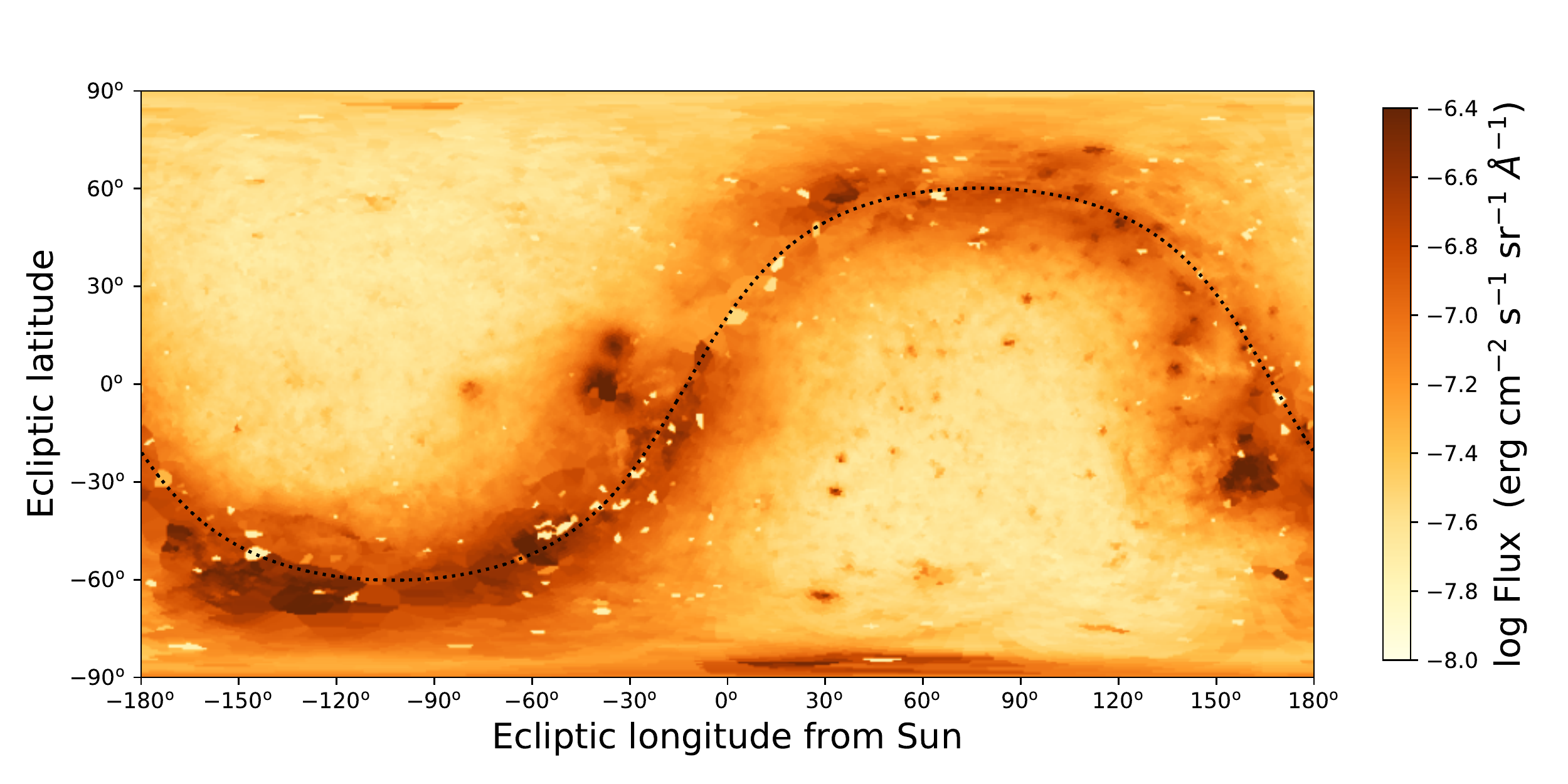}
  \includegraphics[width=3.5in]{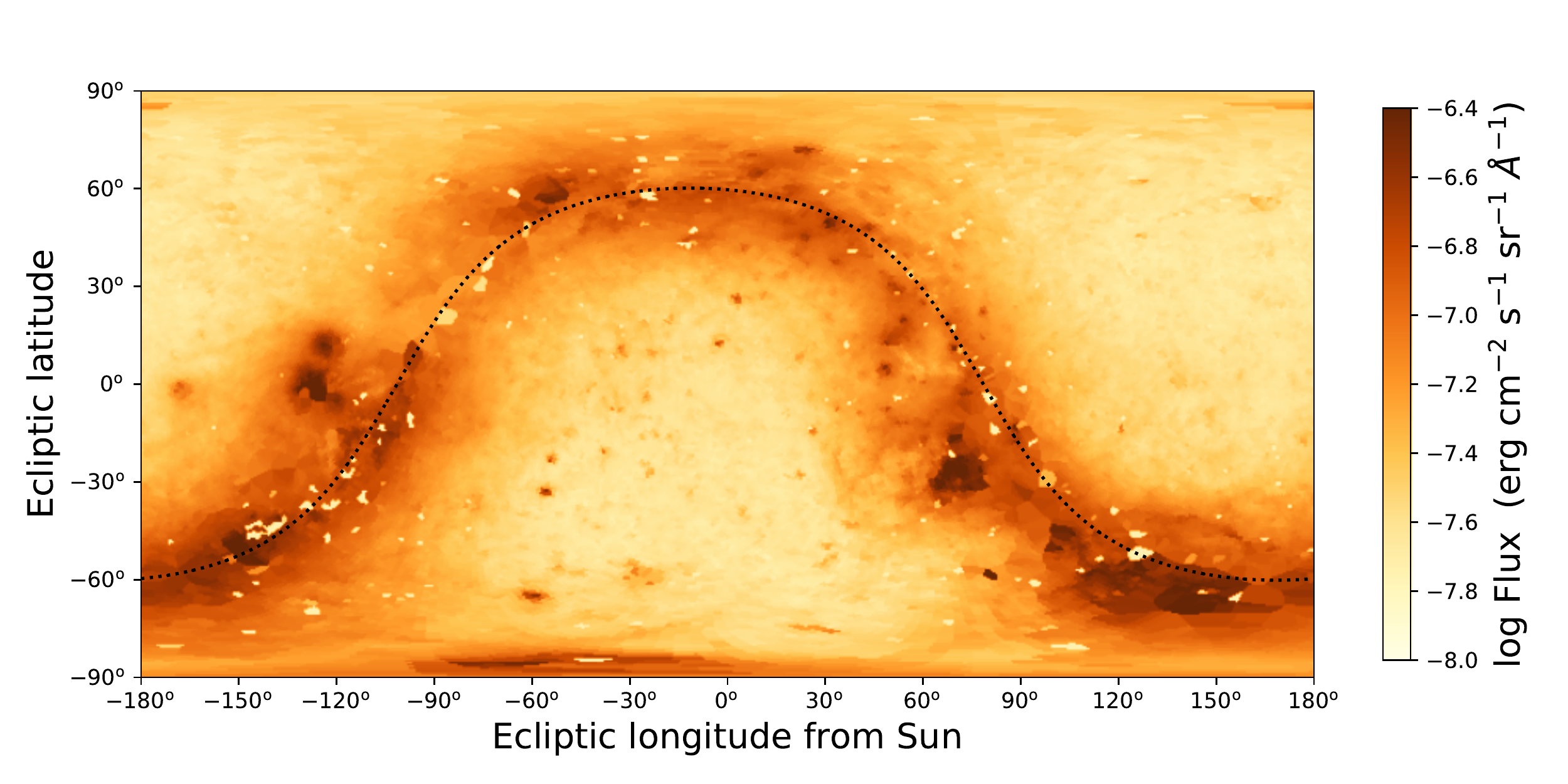}
  \end{center}
  \caption{Total intensity within a near-UV bandpass for the Galactic background,
  based on observations by GALEX.
Two dates are shown: Jan 1 (left) and Apr 1 (right).
In our Sun-centered reference frame, the Galactic plane (dotted curve) shifts from right to left across the sky during the year, while the Zodiacal Cloud remains fixed.
}\label{galmodel}
\end{figure}

Both the overall background flux and its dispersion increase toward the Galactic midplane.
To fully account for the effect of this Galactic structure on observations of the Zodiacal Cloud, we include all of the observed structure directly into our simulations.
We condense the \citet{Murthy2014} catalog into 1-deg$^2$ bins, resulting in an all-sky map of the Galactic background with the same angular resolution as our simulated observations.
Note that as the Earth orbits the Sun, the Galactic contribution rotates behind the Zodiacal dust;
Figure \ref{galmodel} shows the Galactic background at two epochs -- Jan 1 and Apr 1 --
illustrating this shift.
Repeating observations of the same heliocentric fields over the course of 6 months, as described in the following section, makes it possible to disentangle the two contributions, greatly reducing the deleterious effects of Galactic contamination.

\begin{figure}\begin{center}
  \includegraphics[width=7.5in]{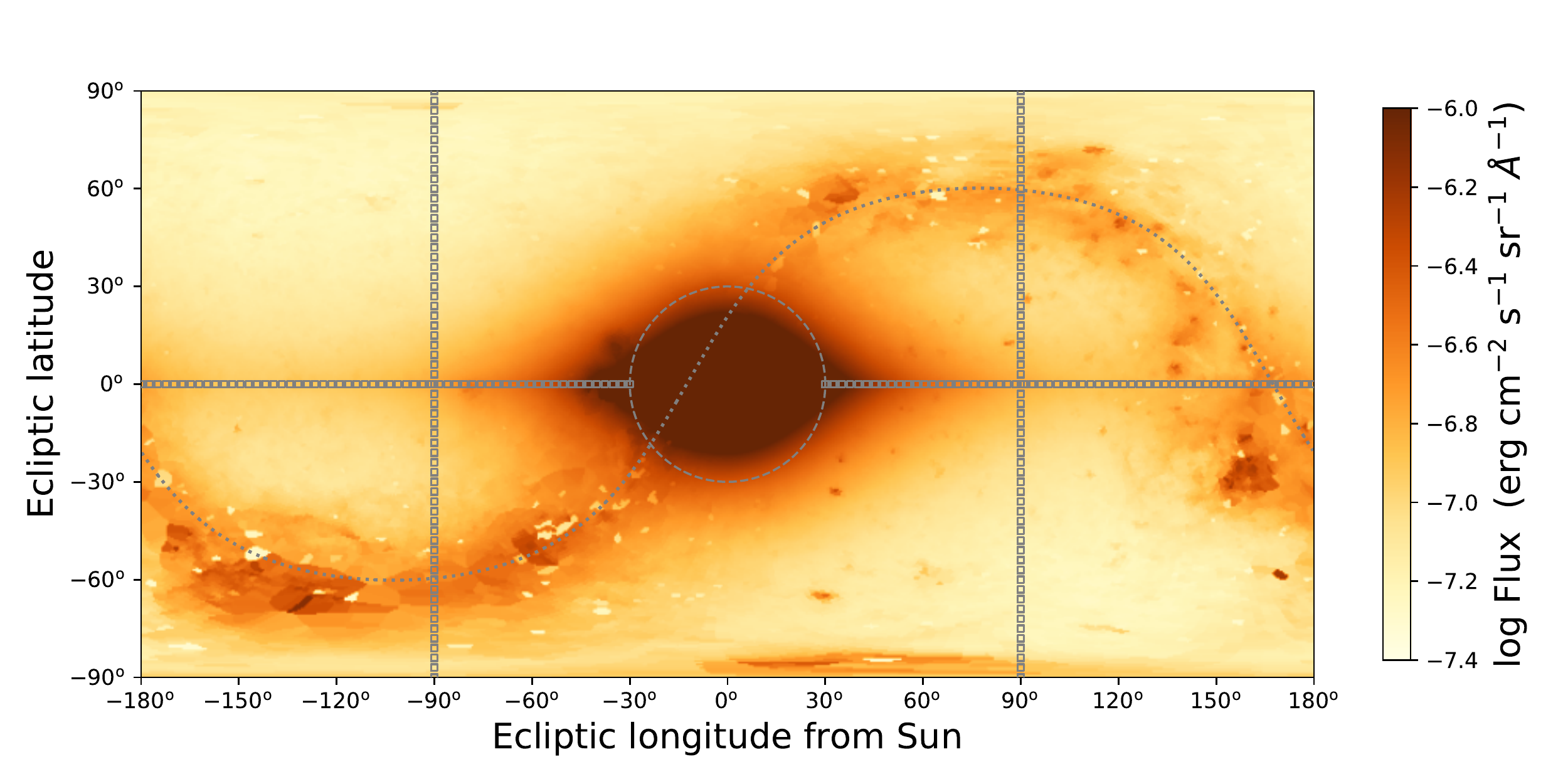} 
  \includegraphics[width=7.5in]{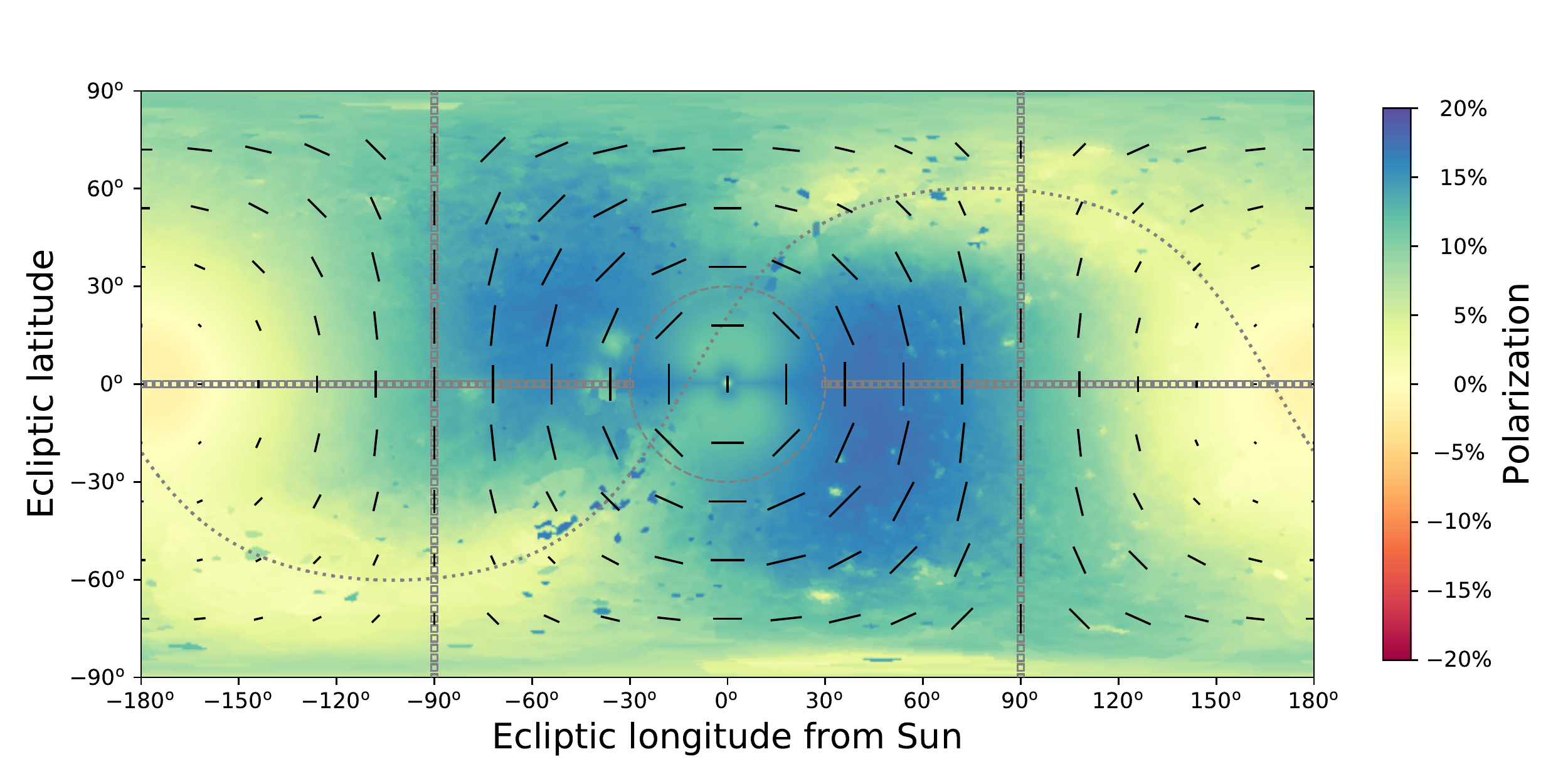}
  \end{center}
  \caption{Total intensity (top) and polarization (bottom) within a near-UV bandpass,
  for our full model, including Galactic background contamination based on observations by GALEX.
  The horizontal rows of squares along the ecliptic plane and vertical rows of squares along a meridian 90\degree\ from the Sun show the locations of simulated observations (\S\ref{observations}), where each square is a 3\degree x3\degree\ field of view. 
Observations are excluded within 30\degree\ of the Sun (dashed circle).
In this Sun-centered reference frame the Galactic plane (dotted curve) shifts from right to left across the sky during the year, while the Zodiacal Cloud remains fixed;
this image corresponds to observations made on Jan 1.
}\label{diskmodel}
\end{figure}

Figure \ref{diskmodel} shows the flux and polarization for our combined model, including both zodiacal dust and Galactic background.

\section{Simulated Observations}\label{observations}

Our intention is to mimic a set of observations that could realistically be obtained with a small space telescope such as the proposed Polarized Zodiacal Light Experiment
\citep[PoZoLE;][]{Turner2022}.
While small satellites cannot compete with flagship missions in terms of angular resolution and sensitivity to point sources, they are well-suited to diffuse light measurements such as those considered here, where the large pixels and large field of view compensate for the smaller collecting area.

\begin{deluxetable}{l|c}
\tablecaption{Assumed Mission Parameters
\label{missionParamTable}}
\tablehead{Parameter & Assumed Performance}
\startdata
Simultaneous bandpasses & 2 \\
UV bandpass & 267 -- 293 nm \\
UV throughput & 10\% \\
Optical bandpass ($B$ band) & 380 -- 490 nm \\
Optical throughput & 25\% \\
\hline
Mission lifetime       & 6 months \\
Number of observing epochs & 10 \\
Number of pointings per epoch & 220 \\
\hline
CCD pixels & 1024 x 1024 \\
Field of view & 3\degree\ x 3\degree \\
Patch size & 1\degree\ x 1\degree \\
Flux precision   & 1\% per patch \\
Polarization precision & 0.01 per patch\\
\hline
Solar exclusion angle  & 30\degree \\
\hline
\enddata
\tablecomments{Throughputs include losses from the waveband filter, a dual-band dichroic, the polarization splitter, and the detector efficiency.}
\end{deluxetable}

We assume a 3\degree\ x 3\degree\ field of view, broken down in 1\degree\ x 1\degree\ patches,
corresponding to $\sim$10\arcsec\ pixels on a 1K$\times$1K CCD.
This large field of view maximizes the photons collected while offering sufficient resolution to mask out nearby UV-bright stars
(GALEX found about $\sim$200 stars per square degree with NUV magnitudes from 12 to 19).
Infrared and optical surveys anyway show little structure in the ZC on smaller angular scales \citep{kelsall98, Koschny2019}.

One pointing is made for each Earth orbit, with the $\sim$45-minute integration time split between two orthogonal polarization angles aligned with the scattering plane.
Uniquely determining the linear polarization generally requires measurements at three angles, but here two are enough since the zodiacal light’s polarization direction is known from the scattering geometry \citep{Koschny2019}.
The minimum surface brightness of the zodiacal light (at the ecliptic pole) is $\nu I_{\nu}$ = 42 nW m$^{-2}$ Sr$^{-1}$
in the NUV band and 315 nW m$^{-2}$ Sr$^{-1}$ in B band \citep{leinert98}. 
For our chosen UV bandpass (267-293 nm) and instrument throughput (10\%, including losses from the waveband filter, a dichroic, the polarization splitter, and the detector efficiency),
this surface brightness can be detected with signal-to-noise ratio (SNR) of hundreds for each polarization direction.
We conservatively assume a SNR of 100 for the flux measurement in each patch, enabling the polarization fraction to be measured to within 0.01.

The focus of these observations is the overall distribution of dust as a function of orbital radius and height above the ecliptic, rather than the narrow bands of dust associated with asteroid families \citep{grogan01}. 
It is therefore sufficient to observe fields along two great circles, one on the ecliptic plane probing radial gradients and the other along the meridian 90\degree\ from the Sun probing gradients out of the plane.
Each pointing along the two circles is separated by the 3\degree\ field of view, resulting in 120 observations along the vertical meridian and another 100 observations around the ecliptic plan, after excluding the region within 30\degree\ of the Sun.

It is critical that the observations extend over several months, such that the Galactic background shifts in our heliocentric-coordinates frame and can then be subtracted out; we assume a 6 month mission lifetime. 
This duration allows 13~repeats of the 220-pointing observing pattern (at one Earth orbit per pointing); we conservatively 
assume 10~such observing epochs spaced over 6~months.
If it is possible to observe twice per orbit around the Earth, flipping the telescope 180\degree{} every half orbit, then twice as many epochs would be attainable.

While the focus of this paper is on UV observations, it would be possible to use the same observatory to measure the Zodiacal Cloud at optical wavelengths.  We have included the UV throughput loss due to a dual-band dichroic in our SNR estimate, but otherwise to not consider optical observations in our modeling effort.
Combining UV with optical would provide a dust color along each line-of-sight, which could assist in associating the dust with various source populations.
Lab measurements of dust grains from the stony S-type asteroid Itokawa \cite{Park2015} and carbonaceous asteroid Ryugu \cite{Nakamura2022}, for example, show compositions consistent with the parent bodies' remotely-sensed colors \cite{Lederer2005,Tatsumi2020}.

Table~\ref{missionParamTable} summarizes the key parameters we adopt for the nominal observing platform.
Based on these assumptions, we calculate a complete set of synthetic observations, apply random photometric noise, and quantify how well the disk parameters are retrieved via a Markov chain Monte Carlo method
\citep[emcee;][]{foreman-mackey13}.  The results are  described in the following section.

\section{Results}\label{results}

The goal of this section is to quantify how well we can extract the underlying disk parameters from the simulated observations.
This is mainly a straightforward Markov chain Monte Carlo (MCMC) analysis, with the same forward model (\S\ref{sec:zcmodel},\S\ref{sec:galBack}) used both to create the observations and to compute likelihoods within the MCMC sampler.
Our prior information is uninformative (flat, uniform), but bounds are applied where quantities that cannot physically go below zero are forced to be positive.  The likelihood function is a standard $\chi^2$ integral of data-model offsets (data minus model divided by uncertainty, squared).

The Galactic background, however, requires special consideration.
We fit the smooth part of the background with four parameters,
as described in \S\ref{sec:galBack} --
the minimum at the Galactic pole,
the slope increasing toward the midplane,
the offset from the midplane, and
a cap on the midplane maximum.
On top of this overall trend there is Galactic substructure -- the main source of noise in our fits.
While this noise is non-Gaussian, potentially introducing bias to the fit, we find that the majority of the scatter in the background can be adequately described by a symmetric 1-$\sigma$ uncertainty as a function of Galactic latitude.  We gauge these uncertainties by the dispersion in our GALEX-derived background map (1-$\sigma$ ranges shown as dotted lines in Figure~\ref{fig:galBack}).
Rather than throw out the regions most contaminated by Galactic structure (e.g.\ everything within 20\degree\ of the Galactic midplane),
we include all observations in our likelihood function to test the robustness of our fitting procedure.

\begin{deluxetable}{cc|c|cc}
\tablecaption{Parameter Fitting Results
\label{fitResultsTable}}
\tablehead{Parameter & Description & Input Value & Retrieved Value & Uncertainty}
\startdata
$\tau_{\rm comet} $ & JFC optical depth & $5.6 \times 10^{-8}$  & $5.3 \times 10^{-8}$ & $0.3 \times 10^{-8}$ \\
$\tau_{\rm asteroid}$ & asteroid optical depth & $1.6 \times 10^{-8}$ & $1.4 \times 10^{-8}$ & $0.2 \times 10^{-8}$\\
$\rho_{\rm oort}$ & Oort-cloud dust density & 5.6 $\times 10^{-9}$ & 5.4 $\times 10^{-9}$ & 0.6 $\times 10^{-9}$ \\
$H/R_{\rm comet}$ & JFC disk height & 0.25 &  0.25 & 0.01 \\
$H/R_{\rm asteroid}$ & asteroid disk height & 0.15 & 0.11 & 0.03 \\
$\gamma_{\rm comet}$ & JFC $\tau$ radial exponent  & $-$0.34 & $-$0.34 & 0.01 \\
$\gamma_{\rm asteroid}$ & asteroid $\tau$ radial exponent & $-$0.34 & $-$0.33 & 0.02 \\
$s_{\rm oort}$ & $\rho_{\rm oort}$ radial exponent & $-$2.5 & $-$2.4 & 0.1 \\
$R_{\rm out}$ & outer edge & 3.3 au & 3.2 au & 0.1 au \\
\hline
$A$ & dust albedo & 0.38 & 0.41 & 0.02 \\
$g_{\rm scat,}\perp$ & perp.\ scattering coeff.\ & 0.68 & 0.68 & 0.01 \\
$g_{\rm scat,}\parallel$ & para.\ scattering coeff.\ & 0.72 & 0.72 & 0.01 \\
\hline
$F_0$ & background at Gal.\ pole & 630 & 628 & 3 \\
$F_{\rm slope}$ & background slope & 500 & 491 & 6 \\
$\ell_{\rm offset}$ & shift in background & $-$3\degree\ & $-$2.9\degree\  & 0.1\degree\ \\
$C_0$ & background cap & 0.13 & 0.13 & 0.01 \\
\enddata
\tablecomments{The three groups of parameters are described in \S\ref{dustDistribution} (dust spatial distribution),
\S\ref{dustscat} (dust scattering properties),
and
\S\ref{sec:galBack} (Galactic background).}
\end{deluxetable}

\begin{figure}\begin{center}
  \includegraphics[width=7.in]{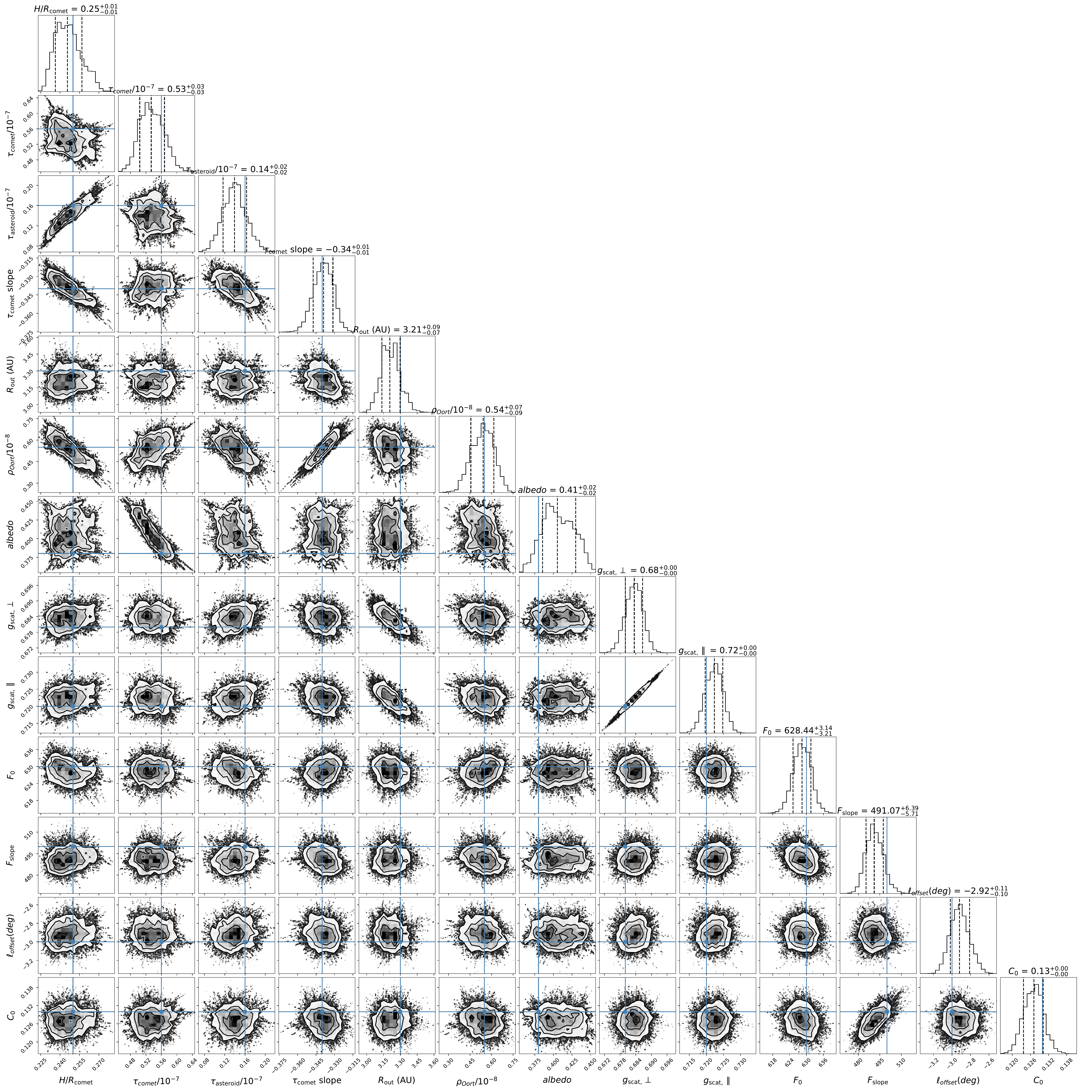}
  \end{center}
  \caption{Posterior distributions are shown for each parameter (upper histograms) with the 1-$\sigma$ ranges marked by dotted lines and compared against the input value (blue lines).  While most parameters exhibit independent, Gaussian-like distributions, cross-correlations reveal some covariance, e.g.\ the two scattering coefficients whose ratio determines the level of polarization ($g_{\rm scat,}\perp$ and $g_{\rm scat,}\parallel$) are highly correlated.
  }\label{mcmcCorner}
\end{figure}

Overall we find that the methodology works, retrieving the key Zodiacal Cloud parameters to sufficient accuracy to disentangle the contributions from the various dust source body populations.
The precisions achieved for the model parameters are listed in Table~\ref{fitResultsTable},
with the full set of parameter cross-correlations is shown in Figure \ref{mcmcCorner}.

\begin{figure}\begin{center}
  \includegraphics[width=4.5in]{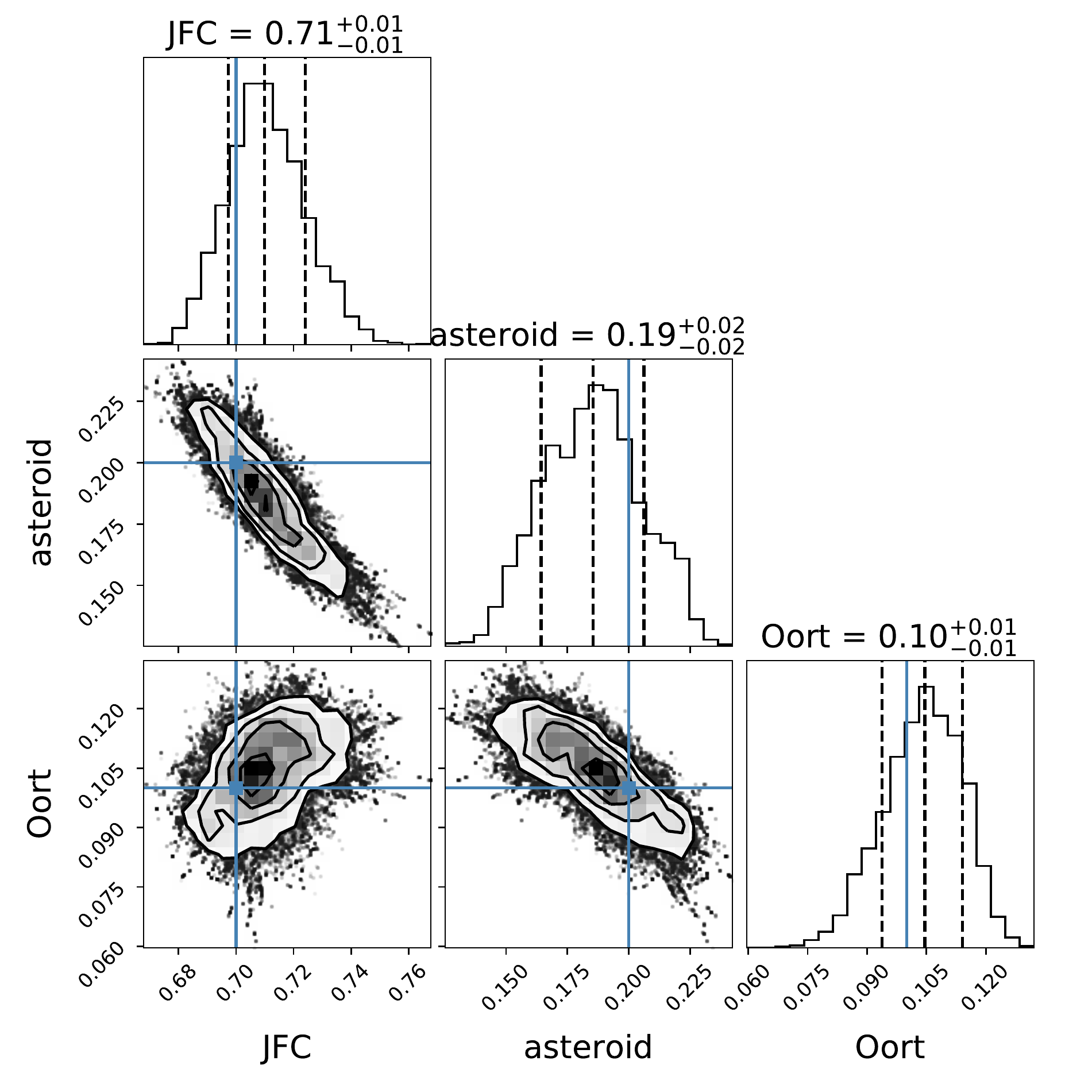}
  \end{center}
  \caption{
    Our primary goal is to measure the optical depth of three contributors to zodiacal dust at 1 au -- Jupiter-family comets, asteroids, and Oort-cloud comets.  The correlations between the retrieved values are shown here.
    Starting with 70/20/10\% allocations for the three components (blue lines show these ``truth" values), each value is retrieved to within 2 percentage point of the total.  The Oort cloud component (the most tenuous contributor) is detected with SNR=10.
       }\label{cornerZodi}
\end{figure}

We comment here on individual parameters of interest:  

\begin{itemize}
\item{{\it Three components of the Zodiacal Cloud:}
The most critical measurement is the relative amounts of dust from the three families of sources bodies.
Figure \ref{cornerZodi} shows the posterior distributions and correlations between the fractional abundances of the three types of dust.
As expected, there is some inverse correlation between the Jupiter-Family comet and asteroidal components (the two that are close to the ecliptic plane), as there is the potential for some interchange between the two densities whil still matching the total column density.
More surprising is that there a similar relationship between the Oort-cloud and asteroidal components.
Most importantly, most difficult component to measure -- the spherically symmetric Oort Cloud component -- is measured to within 1 percentage point (10$\pm$1\%; SNR=10).}

\item{{\it Height of the two disk components:}}
The aspect ratios of the two disk components are measured with 0.01 and 0.03 precision (for the cometary and asteroid components).  
Recall that the observations are spaced uniformly along and vertical meridian.
The lower precision on the asteroid component suggests that concentrating the measurements closer to the ecliptic midplane might be warranted.

\item{{\it Radial exponent to the disk:}}
Repeat observations in the ecliptic plane result in very good measurement of the radial gradient of the disk surface density. 
The exponent to the radial power-law (nominally -0.34) is measured to within 0.02.

\item{{\it Outer edge:}
While the outer edge of the Zodiacal Cloud is also measured with quite good precision (0.1 au), this could be 
improved if future modeling efforts included
in situ radiometry or particle counts by missions traveling to the outer solar system \citep[e.g.][]{Mann2006},
thereby providing a strong prior constraint on the outer edge.
}
\end{itemize}

\begin{figure}\begin{center}
    \includegraphics[width=3.5in]
    {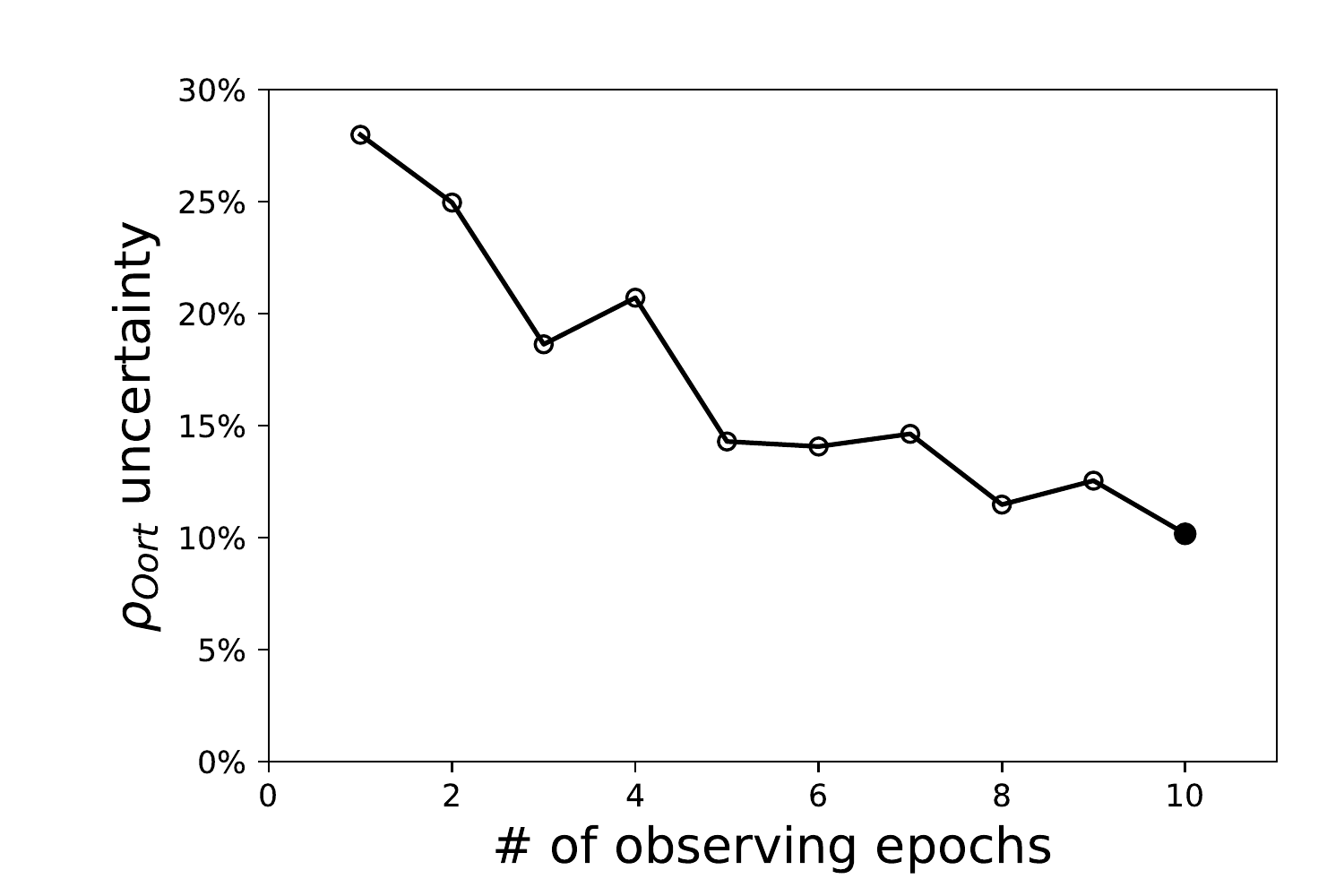}
  \end{center}
 \vspace{-0.2in}
  \caption{Our ability to measure model parameters depends on the input mission parameters, such as the number of observing epochs.
  As one would expect, the uncertainty in the Oort-cloud dust density ($\rho_{\rm Oort}$) improves as more observations are made.
  Our nominal mission has 10 epochs (solid circle).
  }\label{epochsTrend}
\end{figure}

\begin{figure}\begin{center}
     \includegraphics[width=3.5in]
      {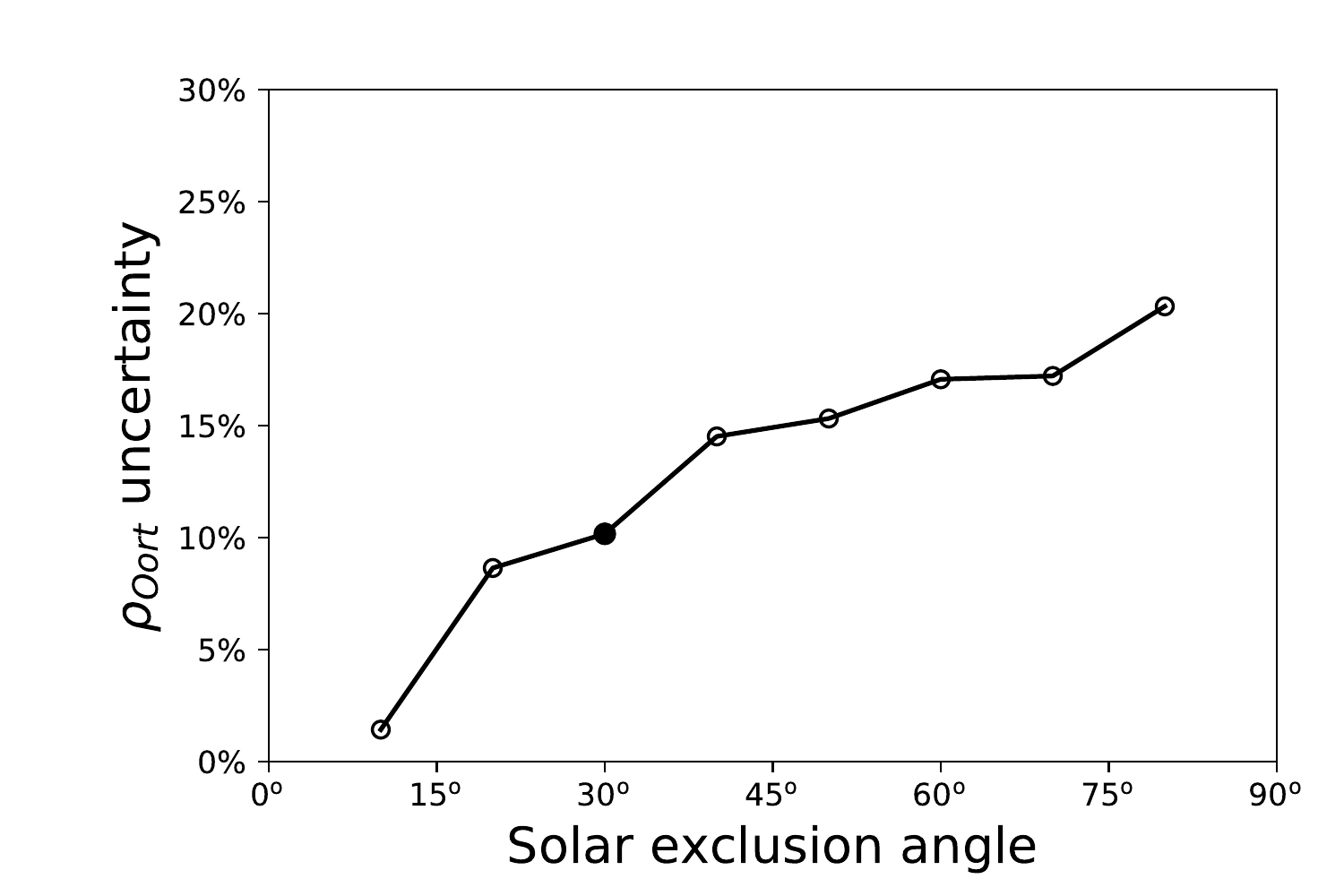}
    \includegraphics[width=3.5in]
    {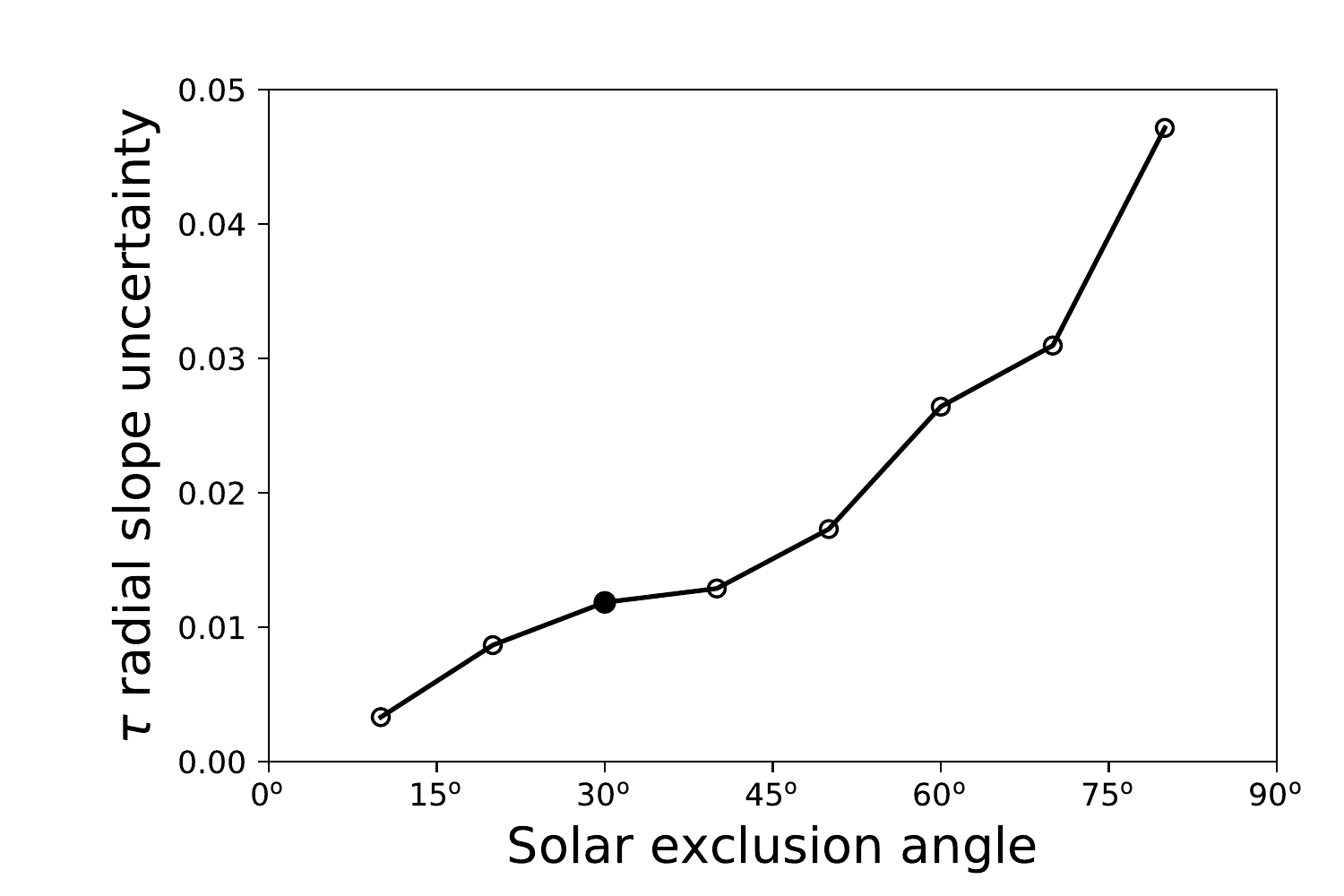}
  \end{center}
 \vspace{-0.2in}
  \caption{Constraints placed on two mode parameters -- $\rho_{\rm Oort}$ and the exponent of the radial power-law -- 
  are shown as a function of the number of the solar exclusion angle.  While most parameters (including $\rho_{\rm Oort}$) have a relatively mild dependence on this angle, the radial exponent is uniquely sensitive to it.
  Our nominal mission has a solar exclusion angle of 30\degree\ (solid circles).
  }\label{solarangTrend}
\end{figure}

The specifics of the above results are of course dependent on the assumed mission parameters 
(Table \ref{missionParamTable}).
A shorter mission with fewer observing epochs will not perform as well, both because of less overall integration time but also because multiple epochs are needed for the Galactic background to rotate relative to the Zodiacal Cloud. 
To test this, we have performed a range of trade studies,
measuring the response as we vary individual parameters.
As expected, changing the measurement SNR 
gives a more or less linear change in the retrieval uncertainties,
while changing the number of observations gives a square-root(N) dependence.
Figure \ref{epochsTrend} shows how the precision on the Oort-cloud dust density changes depending on the number of observing epochs, against consistent with 
a simple square-root(N) dependence.
While most mission parameters have a similar effect on each of the retrieved model parameter uncertainties, 
the solar exclusion angle is an exception.
Excluding a large zone (while keeping the total number of observations constant) decreases the precision of all measurements, but usually with a rather flat dependency, e.g.\
increasing the nomimal angle from 30\degree\ to 80\degree\ only doubles the uncertainty for $\rho_{\rm Oort}$ 
(left panel of Figure \ref{solarangTrend}) 
and most other model parameters.
The solar exclusion angle, however, has a very strong impact on our ability to determine the radial distribution of dust.
For the same change in angle (from 30\degree\ to 80\degree) the uncertainty   on the radial power-law exponent increases by nearly a factor of 5 (right panel of Figure \ref{solarangTrend}) when we can no longer track the material close to the Sun.
  
\section{Summary \& Discussion}\label{summary}

We find that the sky coverage and photometric precision from our nominal mission are sufficient to determine the Zodiacal Cloud's size and shape, as we have demonstrate by synthetically observing radiative transfer models of the cloud and recovering the models' input parameters using a Markov-chain Monte Carlo analysis. The cloud's shape is governed by the orbital inclinations of the component dust particles and so measuring the shape enables distinguishing asteroidal from short-period cometary sources and determining whether Oort Cloud comets contribute a nearly spherically-symmetric dust component.

Revisiting the same heliocentric-coordinates fields after the Milky Way has rotated behind the zodiacal light provides an effective strategy for mitigating the effect of Galactic background light on Zodiacal Cloud observations.
When we synthetically observe the same heliocentric field 10 times spanning at least 6 months,
a tenuous spherically-symmetric (Oort Cloud) component to the dust is clearly detected above the Galactic backgrounds.
NASA’s premier UV observatory, the Hubble Space Telescope, can measure the zodiacal light \citep{Kawara2017} but cannot feasibly revisit enough of the sky over multiple epochs to precisely remove the Galactic background.  Furthermore, Hubble cannot map in polarized UV light to best gauge the cloud’s size and shape, nor can Hubble safely look as close to the Sun as needed to determine the cloud’s appearance inside 1 au.  

While the immediate application of the observations considered here would be to further our understanding of the origin and evolution of our own planetary system,
the results would also have implications for observations of planetary systems around other stars, in particular for a flagship mission to directly image Earth-like planets.
NASA preparation for a such mission involves estimating the frequency of Earth-size planets 
\citep[from Kepler;][]{Bryson2021} and measuring the expected noise level from exo-Zodiacal dust \citep[from LBTI;][]{ertel20}.
The analogous dust clouds veiling target exoplanets may be brighter than our own; the Large Binocular Telescope Interferometer (LBTI) survey of nearby stars found a median exozodiacal brightness of 3 times our Zodiacal Cloud’s, though with large uncertainty \citep{ertel20}.
Further characterization of exozodiacal dust is needed, to assist in interpreting observations of candidate exo-Earths.
There are three specific issues that the observations considered in this paper can address:
1) the color of the dust at UV wavelengths (related to composition),
2) the likelihood of a forward-scattering halo inducing ``pseudo-zodi'' noise (related to dust geometry and scattering properties),
and
3) the capture efficiency of Earth-mimicking dust blobs (related to the orbits of the dust particles).

The first issue (UV characteristics of the dust) arises because of the importance of ozone.
Detection of an Earth-like planet requires a signature of potential habitability.
Among the strongest such signatures is the Hartley absorption band of ozone (O$_3$) at near-UV wavelengths. The Hartley band is so strong it produces a spectral feature even when atmospheric oxygen (O$_2$) levels are too low for detecting signs of O$_2$ itself \citep{TheLUVOIRTeam2019}. 
The band is strong enough to pick out at modest spectral resolutions \citep[R $\simeq$ 6;][]{Gaudi2020} and will be among the first signatures within reach as our technical capabilities advance. 
Ozone’s importance means that a four-meter-class space telescope without a near-UV coronagraph would fail to measure the Earth-like fraction among exoearth candidates \citep{Checlair2021}.
Even with a near-UV coronagraph, detection of ozone depends on the UV albedo of the dust,
with this noise source potentially setting the required size of the telescope \citep{Stark2019}.
The UV described in this paper would provide a direct measure of UV-optical color of dust explicitly linked to source bodies in the solar system and could thereby serve as a reference point for predictions of exozodiacal dust emission at UV wavelengths.

The second issue concerns the so-called ``pseudo-zodi" --
an effect where dust in the outer regions of a planetary system can be brighter than dust in the habitable zone \citep{stark15}.  
This noise source is not due to the amount of outer dust, but rather its geometry (disk-like, edge-on) and its scattering properties (strongly forward scattering).
Improved understanding of the distribution and optical properties
of dust in the solar system will help address the frequency of pseudo-zodi effects around nearby stars, currently estimated at 50\% as common as standard exozodiacal emission \citep{stark15}. 

Lastly there is the third issue of dust clumps in exozodiacal disks, which might be mistaken for planets.
While clumps can arise from recent collisions, the timescale for removal is relatively short, sometimes disappearing within a year \citep{melis12}.
The greater concern for observations of mature planetary systems is the presence of steady-state dust blobs created by the planets themselves as they capture inspiraling dust.  The Earth, for example, induces a $\sim$10\% enhancement of surface density in the dust trailing its orbit \citep{kelsall98}.
Depending on the baseline level of dust and planet mass, such enhancements can emit more flux than the planet itself \citep{stark08}. 
The difficulty in interpreting any observed blobs comes from 
not knowing the orbital elements of the dust; dust on near circular orbits is captured more efficiently than eccentric particles.
So while improvements in models of dust evolution \citep[e.g.\ inclusion of dust collisions;][]{Stark2009} have improved their fidelity, 
not knowing the source bodies for the dust leads to an inherent uncertainty in the shape and optical depth of each simulated dust blob.
Measurements of our Zodiacal Cloud’s structure can be used to validate models of dust sources and transport into the solar system's habitable zone, so that the models become more reliable when applied to exozodiacal clouds far away.


\acknowledgments
Part of this work was carried out at the Jet Propulsion Laboratory, California Institute of Technology, under a contract with the National Aeronautics and Space Administration (80NM0018D0004). P.P.\ was supported by the NASA Solar System Workings award No.\ 80NSSC21K0153. \copyright 2023. 
All rights reserved.

\software{
\texttt{astropy} \citep{astropy2022},
\texttt{emcee} \citep{foreman-mackey13}
}

\footnotesize

\end{document}